# Observing and controlling a Tamm Plasmon at the interface with a metasurface

Oleksandr Buchnev,[1] Alexandr Belosludtsev,[2] Victor Reshetnyak,[3] Dean R. Evans,[4] and Vassili A. Fedotov[1,*]

[1]*Optoelectronics Research Centre and Centre for Photonic Metamaterials, University of Southampton, SO17 1BJ, UK*

[2]*Optical Coating Laboratory, Center for Physical Sciences and Technology, Vilnius LT-02300, Lithuania*

[3]*Physics Faculty, Taras Shevchenko National University of Kyiv, 01601, Ukraine*

[4]*Air Force Research Laboratory, Materials and Manufacturing Directorate, Wright-Patterson Air Force Base, Ohio, USA*

* email: vaf@orc.soton.ac.uk

**Abstract:** We demonstrate experimentally that Tamm plasmons in the near-IR can be supported by a dielectric mirror interfaced with a metasurface, a discontinuous thin metal film periodically patterned on the sub-wavelength scale. More crucially, not only do Tamm plasmons survive the nano-patterning of the metal film, but they also become sensitive to external perturbations, as a result. In particular, by depositing a nematic liquid crystal on the outer side of the metasurface we were able to red-shift the spectral position of Tamm plasmon by 35 nm, while electrical switching of the liquid crystal enabled us to tune the wavelength of this notoriously inert excitation within a 10 nm range.

**Introduction**

A Tamm plasmon (TP) is a localized resonant optical state, a quasi-particle, which exists at the interface between a metal and a dielectric (or semiconductor) Bragg mirror. It was theoretically predicted in [1] and experimentally observed in [2]. The TP dispersion lies completely within the light cone and therefore, in contrast to an ordinary surface plasmon polariton, a Tamm plasmon can be excited with both TE- and TM-polarized light at any angle of incidence [1]. Another advantage of a Tamm plasmon over a surface plasmon polariton is that the former appears to be almost insensitive to dissipative losses in the metal film since its electromagnetic fields are localized predominantly in the non-absorbing Bragg mirror [3]. Because of its robust nature, a Tamm plasmon has been regarded as a viable alternative to conventional surface plasmons in a wide range of applications, including optical switches, semiconductor lasers and light emitters, temperature and refractive-index sensors [4-12]. For many practical applications it is important to realize an external dynamic control over the TP wavelength. Such a task, however, presents a formidable challenge since the fields of a Tamm plasmon reside inside the Bragg mirror and, therefore, are not accessible from the outside. Correspondingly, the approaches proposed so far involve the integration of a control element into the very structure of the Bragg mirror [4,7,13-16], which is not always feasible. It has also been shown that the wavelength of a Tamm plasmon could change (although irreversibly) as a result of either patterning the metal film on the microscale [3,17,18] or corrugating it on the nanoscale [19].

In this Letter, we report on the first experimental observation of a near-IR Tamm plasmon at the interface between a Bragg mirror and a nano-patterned metal film acting as a non-diffracting optical metasurface. We also found that the discrete framework of the metasurface exposed Tamm plasmon to external perturbations, such as changes of the refractive index in an adjacent medium, which enabled us to dynamically control the wavelength of this weakly coupled optical state in a simple yet efficient way.

**Results and Discussion**

Figure 1a presents the design of the structure that was used to observe Tamm plasmons in our experiments. The structure was based on a silver-coated dielectric Bragg mirror designed to exhibit a 0.5 µm wide reflection band centered at the wavelength $\lambda$ = 1.45 µm. It was formed by a stack of alternating 11 layers of $Nb_2O_5$ ($n_{Nb2O5}$ = 2.24 [20]) and 10 layers of $SiO_2$ ($n_{SiO2}$ = 1.47 [21]). The niobium pentoxide and silicone dioxide layers had the thickness of correspondingly 159 ± 2 nm and 246 ± 2 nm, and were deposited onto a double-side polished fused silica substrate using magnetron sputtering (Kurt J. Lesker PVD 225), as detailed in [20]. The silver coating had the thickness of 37 ± 2 nm, and was applied to a section of the Bragg mirror by magnetron sputtering at room temperature, working pressure of 2.2 mTorr and deposition rate of about 11 nm / min using an Ag planar sputtering target (99.99% purity). A 30 µm x 30 µm patch of the silver film was turned into a metallic metasurface by nano-patterning the film with a focused ion beam (Helios Nanolab 600). The pattern of the metasurface featured a square array of 550 nm large disks with the period of 600 nm (see Fig. 1b), which rendered the nanostructure as non-diffracting above $\lambda$ = 1.34 µm, i.e., well within the reflection band of the Bragg mirror (for normally incident light the first diffraction order appeared in the adjacent $Nb_2O_5$ layer at wavelengths shorter than $n_{Nb2O5} \cdot 0.6$ µm). Such a disk pattern, although very simple, was sufficient to ensure that the array would act as a narrow-band reflector – one of the most basic functionalities of metasurfaces [22]. The reflection band was centered at a wavelength of ~ 1.65 µm, as defined by the plasmon resonance of silver nano-disks. The resonant nature of our metasurface implied, in particular, that within the reflection band it would appear optically as dense as the unstructured silver film.

The spectral response of the fabricated sample was characterized in reflection at normal incidence using a commercial microspectrophotometer developed by CRAIC Technologies on the basis of a ZEISS Axio microscope. It employed a cooled near-IR CCD array with spectral resolution of 0.8 nm and featured a tungsten-halogen light source equipped with a broadband linear polarizer.

Light was focused onto the sample as well as collected from the metal side using a x15 reflective objective with NA 0.28. The reflectivity spectra were acquired through a 22 μm x 22 μm square aperture installed in the image plane of the microscope.

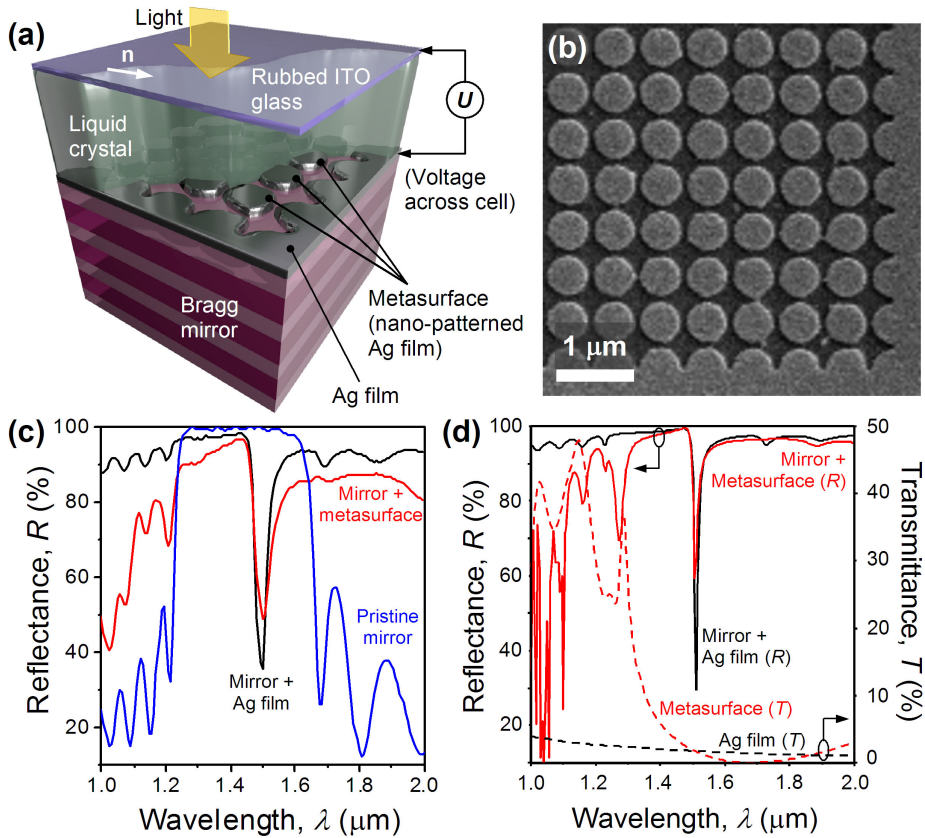

**Fig. 1.** (a) Schematic of the structure used in our experiments. The white arrow indicates the direction of rubbing (applied to underside of ITO cover glass), **n**, which controlled the alignment of the liquid crystal in the cell. (b) Scanning electron micrograph of a fragment of the metasurface fabricated on top of a Bragg mirror. (c) Experimental reflectivity spectra of the Bragg mirror acquired while it was in a pristine state (blue), after it was interfaced with a 37 nm thick continuous silver film (black), and after the silver film was nano-patterned to become a metasurface (red). (d) Calculated spectra of the continuous silver film and metasurface. Solid curves show reflectivity of the film (black) and metasurface (red) placed atop of a Bragg mirror, while dashed curves show transmission of the film (black) and metasurface (red) residing on a niobium pentoxide substrate.

Figure 1c compares the reflectivity spectra taken at three different areas of the sample corresponding to an uncoated (i.e. pristine) Bragg mirror, a Bragg mirror with a continuous silver film and a Bragg mirror with the metasurface. As per design, the pristine mirror is seen to exhibit a characteristic, spectrally flat reflection band spanning from about 1.22 to 1.66 µm (blue curve in Fig. 1c). The reflectivity spectrum of the silver-covered area of the mirror reveals the appearance of a narrow reflectivity dip located within the band of the pristine mirror and centred at $\lambda = 1.49$ µm (black curve in Fig. 1c). Such a conspicuous transformation of the Bragg mirror's reflectivity spectrum signifies the excitation of the Tamm plasmon, as has previously been shown in a number of works [2,23-26]. Intriguingly, the Bragg mirror, when combined with the metasurface, also appeared to support Tamm plasmons, exhibiting a similar reflectivity dip in the same spectral window (red curve in Fig. 1c). While Tamm plasmons have been shown to exist at the interface with micrometer-sized metal patches [3,17,18] and nano-corrugated metal films [19], our results represent, to the best of our knowledge, the first experimental evidence that Tamm plasmons can also survive the segmentation of metal films into nanometer-sized patches.

Such an outcome is quite remarkable in the view that sub-wavelength metal patches promote very strong coupling of optical fields to leaky modes of an uncovered Bragg mirror and, therefore, alone cannot support Tamm plasmons [27]. The key here is to arrange metal patches into an array of sub-wavelength periodicity, which will reduce the leakage to a reasonable level and also make sure that the resulting nano-structure behaves optically as a continuous film (and, thus, can be regarded as a metasurface). Furthermore, within the resonance band of such a metasurface the polarizability of the metal patches is naturally larger than that of an unpatterned metal film, and hence their scattering cross-section can exceed their physical size [28] effectively making up for the lack of metal between the patches. That, in particular, minimized the distinction between our metasurface and continuous silver film at around 1.5 µm, which was why the nano-patterning did not seem to affect the TP wavelength in the experiment.

Our observations are supported by the results of full-wave numerical modeling (see Fig. 1d), which was carried out using commercial simulation software COMSOL Multiphysics. Indeed, the modeled reflectivity spectra of the continuous silver film and metasurface placed atop of the Bragg mirror are seen to exhibit a TP resonance at the same wavelength, $\lambda = 1.51$ μm. The slight mismatch between the calculated and experimentally measured values of the TP wavelength is attributed to a deviation of the dielectric function of deposited silver from the tabulated data used in our simulations [29]. Note that the transmission of the metasurface alone varies strongly with the wavelength (as dictated by the plasmon resonance) and at $\lambda = 1.51$ μm it matches the transmission of a standalone continuous film exactly. The importance of this observation becomes apparent if one recalls that the transmission coefficient of optically thin metal films is almost exclusively controlled by the imaginary part of the refractive index $k$ [30] and, hence, by Re $\varepsilon$ (since for metals $|k| \gg n$ in the near-IR). Given that the reflectivity of the standalone silver film and metasurface at $\lambda = 1.51$ μm differs by less than 2 % (not shown in Fig. 1d), one may expect that their complex dielectric constants at this wavelength will be nearly identical. Indeed, based on the modeled transmission and reflection data we obtained $\varepsilon = -87 - i12$ and $\varepsilon = -88 - i7$ for the metasurface and silver film, respectively.

While the nano-pattering of the silver film did not affect the ability of the structure to support a Tamm plasmon, it naturally exposed the surface of the Bragg mirror. Consequently, that should have locally broken the confinement of the Tamm plasmon, allowing a direct access to its fields (which would otherwise remain difficult to couple to, residing under the continuous silver film [8,31,32]). To verify that assumption experimentally, we introduced an electrically controlled liquid-crystal (LC) cell into the structure of the sample, as schematically shown in Fig. 1a. The cell was assembled by placing an ITO cover glass approximately 10 μm above the silver-coated surface of the mirror. The cover glass served as the top (transparent) electrode of the cell, while the silver film played the role of its bottom electrode. The cell was vacuum-filled with E7 (Merck), a widely used and commercially available LC mixture with high optical anisotropy ($n_o = 1.50$, $n_e = 1.70$ at $\lambda = 1.55$ μm [33]). The surface of the cover glass facing the mirror had been coated with a thin film of uniformly rubbed polyimide to ensure planar alignment of LC molecules in the cell (i.e., parallel to the mirror and along the direction of rubbing, **n**). By increasing the voltage across the cell, $U$, we gradually switched E7 from the planar to homeotropic state in which LC molecules were oriented perpendicular to the mirror. Due to optical anisotropy of LC molecules the switching of the cell was accompanied by the change of the LC refractive index from $n_e$ to $n_o$ for light polarized parallel to the direction of rubbing (**E** ∥ **n**). Correspondingly, for light polarized perpendicular to the direction of rubbing (**E** ⊥ **n**) the refractive index remained $n_o$.

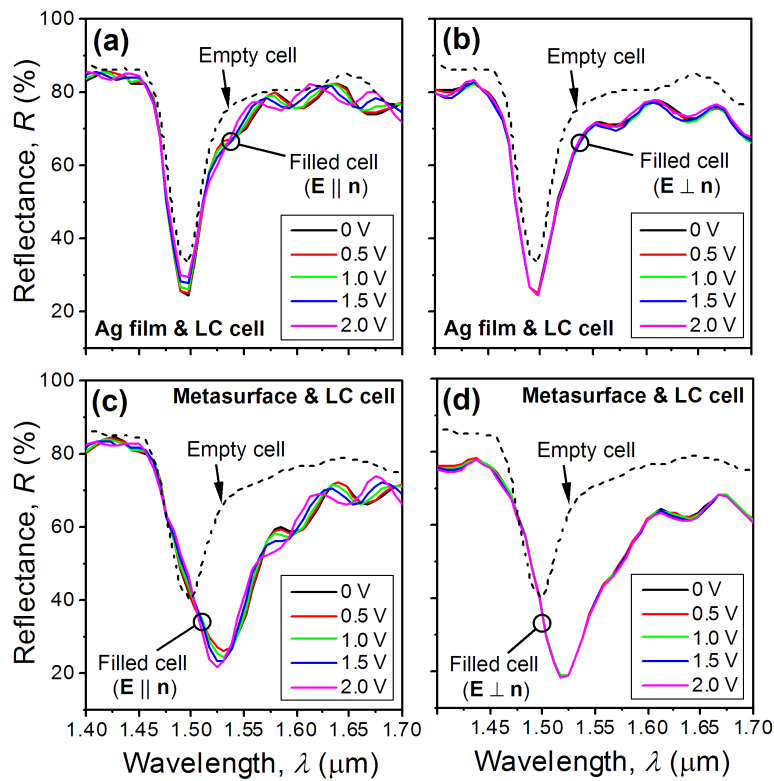

**Fig. 2.** Experimental reflectivity spectra of the silver-coated Bragg mirror acquired with linearly polarized light at different voltages once the structure was integrated with a liquid-crystal cell (solid curves). Data in panels (a) and (b) correspond to an area of unpatterned silver film, while in panels (c) and (d) – to the metasurface. Dashed curves show the reflectivity spectra of the sample before the cell was filled with the liquid crystal.

We note that although the metasurface was electrically discontinuous and, therefore, could not act as the bottom electrode, the section of the cell above it, when inspected under a polarizing microscope, was also displaying the transition between the planar and homeotorpic states. Such a behavior was a testament to the long-range nature of the elastic forces arising in nematic liquid crystals. More specifically, the effective range of the elastic forces in an LC cell is comparable to the thickness of the cell [34], which in our case ensured that the voltage-induced orientation of LC molecules would be transferred from the periphery of the metasurface to most of the area above it.

Figure 2 presents reflectivity spectra of the sample integrated with the LC cell, which were measured under linearly polarized light, while sweeping the voltage across the cell from 0 to 2 V. Quite evidently, in the case of unstructured silver film the spectral location of the Tamm plasmon was unaffected by filling the cell with E7, as well as by changing the state of the liquid crystal in the cell (see Figs. 2a and 2b). As noted above, such behavior resulted from strong confinement of the TP fields, which were effectively screened by the unstructured film from the ambient medium and, expectedly, remained inert to external perturbations [8,31,32]. By contrast, the Tamm plasmon excited at the interface with the metasurface appeared to be quite sensitive to the presence of the liquid crystal (see Figs. 2c and 2d). In particular, filling the cell with E7 red-shifted the TP reflectivity dip by about 35 nm ($\mathbf{E} \parallel \mathbf{n}$) and 25 nm ($\mathbf{E} \perp \mathbf{n}$), which was consistent with the increase of the refractive index above the metasurface from 1 to $n_e$ and $n_o$, respectively. Also, for $\mathbf{n}$-polarized illumination the TP reflectivity dip was seen to blue-shift as soon as the applied voltage had exceeded 0.5 V, as evident from Fig. 2c. The extent of the shift reached 10 nm at $U = 2.0$ V (see Fig. 3a) and corresponded to the change of the refractive index in the LC cell from $n_e$ and $n_o$. Our observations, thus, appeared to agree with the assumption we made earlier that the nano-patterning of the silver film would allow external coupling to the fields of the Tamm plasmon. Note that for the orthogonal polarization ($\mathbf{E} \perp \mathbf{n}$) sweeping the voltage did not have any effect on the TP wavelength (see Figs. 2d and 3a). Indeed, in that case the re-orientation of LC molecules occurred in the plane perpendicular to the polarization of light and, therefore, could not affect the relevant component of the refractive index tensor.

The data set presented in Fig. 2 also confirms that the spectral shift of the TP reflectivity dip could not result from the selective excitation (due to focused illumination) of TE- and TM-polarized Tamm plasmons. Indeed, although TP resonances obtained with TE and TM polarizations do not share the same wavelength [1], their excitation would require TE-TM polarization conversion to occur inside the liquid crystal and, therefore, could not be exclusive to just one particular case, namely when the metasurface was illuminated with **n**-polarized light (Fig. 2c). More crucially, we reproduced the observed behavior by modeling in COMSOL the reflectivity of our sample at normal incidence under plane-wave illumination (see Figs. 3b and 3c). Note that calculated shift of the TP resonance (12 nm) resulting from LC transition between the planar and homeotropic states above the metasurface only marginally exceeds the shift we observed in the experiment, which confirms that the re-orientation of LC molecules above the metasurface was nearly complete.

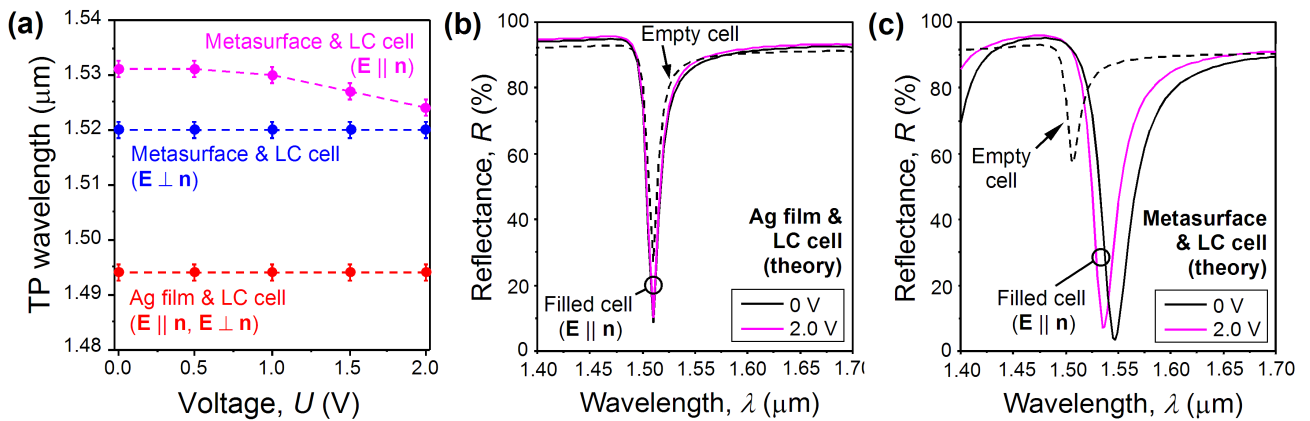

**Fig. 3.** (a) Wavelengths of the Tamm plasmon resonance plotted as functions of applied voltage. Data points were extracted from Figs. 2a - 2c. Dashed curves are a guide for the eye. Panels (b) and (c) display calculated reflectivity spectra of a continuous silver film and metasurface atop of a Bragg mirror integrated with a liquid-crystal cell, respectively. Dashed curves show the reflectivity of the structures before the cell was filled with a liquid crystal. Solid curves correspond to the planar (black) and homeotropic (magenta) states of the liquid crystal.

Apart from changes in the spectral location of the TP reflectivity dip, we observed consistent, voltage-induced changes of its magnitude under **n**-polarized illumination. Interestingly enough, such changes occurred also in the case of unstructured film (see Fig. 2a). The exact nature of those changes remains unclear to us. We speculate that the effect might result from mild focusing of light produced by the objective of our microspectrophotometer. Indeed, for light focused on a metal-coated Bragg mirror a change in the refractive index immediately above the structure would move its image from the image plane of the instrument and could, in principle, affect the level of the measured reflectivity. Another possible explanation is that a change in the refractive index of the ambient medium would change the reflectivity of its interface with the metal film and, therefore, could modify the conditions for the excitation of a Tamm plasmon. Such an explanation appears to be consistent with the results of our numerical simulations presented in Figs. 3b and 3c, suggesting a different approach to sensing with Tamm plasmons. A more detailed analysis of the above noted effect is currently underway and will be published elsewhere.

**Conclusions**

In conclusion, we showed experimentally that a Tamm plasmon could be excited in the near-IR at the interface between a dielectric Bragg mirror and a nano-structured non-diffracting metasurface (which effectively replaced a continuous metal film used conventionally as the second mirror). Our findings also indicate that the metasurface, through its discrete framework, had enabled an external access to, otherwise, weakly coupled fields of the Tamm plasmon. More specifically, we found that placing a dielectric, such as a liquid crystal, in direct contact with the outer side of the metasurface red-shifted the TP wavelength by as much as 35 nm, while no spectral shift could be detected when the liquid crystal was applied to a continuous metal film in the conventional configuration. Furthermore, we managed to tune the TP wavelength within a 10 nm range by changing the LC refractive index above the metasurface with an externally applied electric field. While the reported tuning range is modest compared to what other TP tuning approaches can offer [4,7,13-16], we point out that our work aimed at proof-of-principle demonstration of the new mechanism for controlling Tamm plasmons. There are a number of ways one can improve the efficiency of the new mechanism, which include, for instance, optimization of the patterning and thickness of the metal film, adjusting the composition of the Bragg mirror, implementing other LC switching modes and employing liquid crystals with higher optical birefringence (e.g., some of the currently available LC materials exhibit birefringence as high as 0.8 [35]). Hence, we argue that the demonstrated ability to control the spectral location of Tamm plasmon opens up a viable route to exploiting this resonant optical state in many real-life applications, including optical switching, enhancement of optical nonlinearity, lasing and light emission, and surface-enhanced spectroscopy.

**Acknowledgments**

The authors acknowledge the financial support of the UK's Engineering and Physical Sciences Research Council (EPSRC) EP/R024421/1 and European Office of Aerospace Research and Development (EOARD) 15IOE011.

**References**

1. M. Kaliteevski, I. Iorsh, S. Brand, R. A. Abram, J. M. Chamberlain, A. V. Kavokin, and I. A. Shelykh, "Tamm plasmon-polaritons: Possible electromagnetic states at the interface of a metal and a dielectric Bragg mirror," Phys. Rev. B **76**, 165415 (2007).

2. M. E. Sasin, R. P. Seisyan, M. A. Kalitteevski, S. Brand, R. A. Abram, J. M. Chamberlain, A. Yu. Egorov, A. P. Vasil'ev, V. S. Mikhrin, and A. V. Kavokin, "Tamm plasmon polaritons: Slow and spatially compact light," Appl. Phys. Lett. **92**, 251112 (2008).

3. I. Yu. Chestnov, E. S. Sedov, S. V. Kutrovskaya, A. O. Kucherik, S. M. Arakelian, and A. V. Kavokin, "One-dimensional Tamm plasmons: Spatial confinement, propagation, and polarization properties," Phys. Rev. B **96**, 245309 (2017).

4. W. L. Zhang and S. F. Yu, "Bistable switching using an optical Tamm cavity with a Kerr medium," Opt. Commun. **283**, 2622–2626 (2010).

5. C. Symonds, A. Lemaitre, P. Senellart, M. H. Jomaa, S. Aberra Guebrou, E. Homeyer, G. Brucoli, and J. Bellessa, "Lasing in a hybrid GaAs/silver Tamm structure," Appl. Phys. Lett. **100**, 121122 (2012).

6. C. Symonds, G. Lheureux, J. P. Hugonin, J. J. Greffet, J. Laverdant, G. Brucoli, A. Lemaitre, P. Senellart, and J. Bellessa, "Confined Tamm Plasmon Lasers," Nano Lett. **13**, 3179-3184 (2013).

7. W. L. Zhang, F. Wang, Y. J. Rao, and Y. Jiang, "Novel sensing concept based on optical Tamm plasmon," Opt. Express **22**, 14524–14529 (2014).

8. B. Auguie, M. C. Fuertes, P. C. Angelome, N. L. Abdala, G. J. Soler Illia and A. Fainstein, "Tamm plasmon resonance in mesoporous multilayers: toward a sensing application," ACS Photon. **1**, 775–80 (2014).

9. S. Kumar, P. S. Maji, R. Das, "Tamm-plasmon resonance based temperature sensor in a Ta2O5/SiO2based distributed Bragg reflector," Sens. Actuat. A **260**, 10–15 (2017).

10. Z. Y. Yang, S. Ishii, T. Yokoyama, T. D. Dao, M. G. Sun, P. S. Pankin, I. V. Timofeev, T. Nagao, and K. P. Chen, "Narrowband Wavelength Selective Thermal Emitters by Confined Tamm Plasmon Polaritons," ACS Photon. **4**, 2212−2219 (2017).

11. S.- G. Huang, K.- P. Chen and S.- C. Jeng, "Phase sensitive sensor on Tamm plasmon devices," Opt. Mater. Express **7**, 1267-1273 (2017).

12. A. Jiménez-Solano, J. F. Galisteo-López, H. Míguez, "Flexible and Adaptable Light-Emitting Coatings for Arbitrary Metal Surfaces based on Optical Tamm Mode Coupling," Adv. Opt. Mater. **6**, 1700560 (2018).

13. H. X. Da, Z. Q. Huang, and Z. Y. Li, "Electrically controlled optical Tamm states in magnetophotonic crystal based on nematic liquid crystals," Opt. Lett. **34**, 1693-1695 (2009).

14. J. Luo, P. Xu, and L. Gao, "Controllable switching behavior of optical Tamm state based on nematic liquid crystal," Solid State Commun. **151**, 993-995 (2011).

15. P. S. Pankin, S. Ya. Vetrov, and I. V. Timofeev, "Tunable hybrid Tamm-microcavity states," J. Opt. Soc. Am. B **34**, 2633-2639 (2018).

16. H. C. Cheng, C. Y. Kuo, Y. J. Hung, K. P. Chen, and S. C. Jeng, "Liquid-Crystal Active Tamm-Plasmon Devices," Phys. Rev. Appl. **9**, 064034 (2018).

17. O. Gazzano, S. Michaelis de Vasconcellos, K. Gauthron C. Symonds, J. Bloch, P. Voisin, J. Bellessa, A. Lemaıtre, and P. Senellart, "Evidence for Confined Tamm Plasmon Modes under Metallic Microdisks and Application to the Control of Spontaneous Optical Emission," Phys. Rev. Lett. **107**, 247402 (2011).

18. M. Aams, B. Cemlyn, I. Henning, M. Parker, E. Harbord, and R. Oulton, "Model for confined Tamm plasmon devices," J. Opt. Soc. Am. B **36**, 125-130 (2019).

19. A. R. Gubaydullin, C. Symonds, J.-M. Benoit, L. Ferrier, T. Benyattou, C. Jamois, A. Lemaître, P. Senellart, M. A. Kaliteevski, and J. Bellessa, "Tamm plasmon sub-wavelength structuration for loss reduction and resonance tuning," Appl. Phys. Lett. **111**, 261103 (2017).

20. K. Juškevičius, M. Audronis, A. Subačius, S. Kičas, T. Tolenis, R. Buzelis, R. Drazdys, M. Gaspariūnas, V. Kovalevskij, A. Matthews, A. Leyland, "Fabrication of Nb2O5/SiO2 mixed oxides by reactive magnetron co-sputtering," Thin Solid Film. **589**, 95-104 (2015).

21. L. Gao, F. Lemarchand, and M. Lequime, "Exploitation of multiple incidences spectrometric measurements for thin film reverse engineering," Opt. Express **20**, 15734-15751 (2012).

22. B. A. Munk, *Frequency Selective Surfaces: Theory and Design* (Wiley, New York 2000).

23. K. J. Lee, J. W. Wu, and K. Kim, "Enhanced nonlinear optical effects due to the excitation of optical Tamm plasmon polaritons in one-dimensional photonic crystal structures," Opt. Express **21**, 28817–23 (2013).

24. B. Auguie, A. Bruchhausen and A. Fainstein, "Critical coupling to Tamm plasmons," J. Opt. **17**, 035003 (2015).

25. C.-Y. Chang, Y.-H. Chen, Y.-L. Tsai, H.-C. Kuo, and K.-P. Chen, "Tunability and Optimization of Coupling Efficiency in Tamm Plasmon Modes," IEEE J. Select. Topic. Quant. Electron. **21**, 4600206 (2015).

26. A. Kumari, S. Kumar, M. K. Shukla, G. Kumar, P. S. Maji, R. Vijaya and R. Das, "Coupling to Tamm-plasmon-polaritons: dependence on structural parameters," J. Phys. D: Appl. Phys. **51**, 255103 (2018).

27. I. Yu. Chestnov, E. S. Sedov, S. V. Kutrovskaya, A. O. Kucherik, S. M. Arakelian, and A. V. Kavokin, "One-dimensional Tamm plasmons: Spatial confinement, propagation, and polarization properties," Phys. Rev. B **96**, 245309 (2017).

28. C. F. Bohren and D. R. Huffman, *Absorption and Scattering of Light by Small Particles*, 2nd ed. (Wiley-Interscience, New York, 1998).

29. A. D. Rakić, A. B. Djurišic, J. M. Elazar, and M. L. Majewski, "Optical properties of metallic films for vertical-cavity optoelectronic devices," Appl. Opt. **37**, 5271-5283 (1998).

30. S. G. Tomlin, "Optical reflection and transmission formulae for thin films," J. Phys. D: Appl. Phys. **1**, 1667 (1968).

31. R. Badugu, E. Descrovi, J. R. Lakowicz, "Radiative decay engineering: Tamm state-coupled emission using a hybrid plasmonic photonic structure," Anal. Biochem. **445**, 1–13 (2014).

32. S. Kumar, M. K. Shukla, P. S. Maji, R. Das, "Self-referenced refractive index sensing with hybrid-Tamm-plasmon-polariton modes in sub-wavelength analyte layers," J. Phys. D: Appl. Phys. **50**, 375106 (2017).

33. J. Li, Sh. T. Wu, S. Brugioni, R. Meucci, S. Faetti, "Infrared refractive indices of liquid crystals," J. Appl. Phys. **97**, 073501 (2005).

34. H. P. Hinov, "Penetration depth of surface forces into nematic layers," Mol. Cryst. Liq. Cryst. **74**, 39-53 (1981).

35. Y. Arakawa, S. Kang, H. Tsuji, J. Watanabe, and G.I. Konishi, "The design of liquid crystalline bisolane-based materials with extremely high birefringence," RSC Adv. **6**, 92845-92851 (2016).